\documentclass[%
 reprint,
 amsmath,amssymb,
 aps,
]{revtex4-1}

\usepackage{amssymb,amsmath,amscd}
\usepackage{graphicx}% Include figure files
\usepackage{dcolumn}% Align table columns on decimal point
\usepackage{bm}% bold math
\usepackage[mathlines]{lineno}% Enable numbering of text and display math
\usepackage[utf8]{inputenc}
\usepackage{dsfont}
\usepackage{amssymb,amsmath}
\usepackage{booktabs}
\usepackage{dsfont}
\usepackage{graphicx}

\begin{document}

\newcommand{\qs}{Q_{\rm sat}}
\newcommand{\qsa}{Q_{\rm sat, A}}
\newcommand{\rr}{\mbox{\boldmath $r$}}
\newcommand{\rrn}{\mbox{$r$}} 
\newcommand{\pom}{I\!\!P} 
\newcommand{\rp}{\mbox{\boldmath $p$}} 
\newcommand{\rqq}{\mbox{\boldmath $q$}} 
\newcommand{\lsim}{\raisebox{-0.5mm}{$\stackrel{<}{\scriptstyle{\sim}}$}}
\newcommand{\gsim}{\raisebox{-0.5mm}{$\stackrel{>}{\scriptstyle{\sim}}$}}
\def\simge{\mathrel{%
   \rlap{\raise 0.511ex \hbox{$>$}}{\lower 0.511ex \hbox{$\sim$}}}}
\def\simle{\mathrel{
   \rlap{\raise 0.511ex \hbox{$<$}}{\lower 0.511ex \hbox{$\sim$}}}}

%\preprint{APS/123-QED}

\title{Geometrical scaling description for the exclusive production of vector mesons and DVCS}% Force line breaks with \\
\thanks{An extended version of this work has been published in \cite{ben2017}}%

\author{F. G. Ben}
 \email{felipeben@gmail.com}
\author{M.V.T. Machado}%
 \email{magno.machado@ufrgs.br}
\affiliation{%
High Energy Physics Phenomenology Group, GFPAE IF-UFRGS 
\\Caixa Postal 15051, CEP 91501-970 Porto Alegre, RS, Brazil
}%
\author{W. K. Sauter}%
 \email{werner.sauter@gmail.com}
\affiliation{%
Instituto de F\'{\i}sica e Matem\'atica,  Universidade
Federal de Pelotas\\
Caixa Postal 354, CEP 96010-900, Pelotas, RS, Brazil
}%


%\affiliation{
%\affiliation{
%\affiliation{%


\date{March, 2018}% It is always \today, today,

\begin{abstract}
In this work, we investigate the exclusive production of particles in scattering processes in the so-called saturation region. Within this scheme the phenomenon of geometric scaling takes place: cross sections are functions only of a dimensionless combination of the relevant kinematic variables, which happens both in inclusive and diffractive cases, as in the production of vector mesons. In particular, the scaling variable is given in general by $\tau = Q^2/Q_s^2$, where $Q^2$ is the photon virtuality and $Q_s$ represents the saturation scale, which drives the energy dependence and the corresponding nuclear effects.
Based on the scaling property, we are able to derive a universal expression for the cross sections for the exclusive vector meson production and deeply virtual Compton scattering (DVCS) in both photon-proton and photon-nucleus interactions. This phenomenological result describes all available data from DESY-HERA for $\rho$, $\phi$ and $J/\psi$ production and DVCS measurements. A discussion is also carried out on the size of nuclear shadowing corrections on photon-nucleus interaction.

\end{abstract}

\maketitle

%\tableofcontents

\section{\label{sec:level1}Introduction}

One of the striking consequences of high energy deep inelastic electron-proton (or electron-nucleus) scattering (DIS) is the geometrical scaling phenomenon. In this regime, the total the total
$\gamma^* p$ and  $\gamma^* A$ cross sections are not a function of
the two independent variables $x$ (Bjorken scale) and $Q^2$ (photon virtuality), but are rather a
function \cite{travwaves} of a single scaling variable, $\tau_A = Q^2/Q_{\mathrm{sat,A}}^2$. The saturation scale $Q_{\mathrm{sat,A}}^2(x;\,A)\propto xG_A(x,\, Q_{\mathrm{sat}}^2) /(\pi R_A^2)$, is connected with gluon saturation effects. At very small $x$, the strong rise of the gluon distribution function is expected to be controlled by saturation. It was demonstrated \cite{Iancu:2002tr}, however, that geometric scaling is not confined to the low momenta kinematic region, it is in fact preserved by the QCD evolution up to relative large virtualities. For proton target, it extends up to $Q^2 \sim Q_{\mathrm{sat}}^4 (x)/\Lambda^2_{\mathrm{QCD}}$,  provided one stays in small-$x$ region. For nuclear targets, that kinematic window is further enlarged due to the nuclear enhancement of the saturation scale, $Q_{\mathrm{sat,A}}^2\simeq A^{1/3}Q_{\mathrm{sat,p}}^2$. It was proven for the first time in Ref. \cite{Stasto:2000er} that the DESY-HERA $ep$ collider data
on the proton structure function $F_2$ present a scaling pattern 
at $x \leq 0.01$ and $Q^2 \leq 400 GeV^2$. Similar behavior was further observed on  electron-nuclei processes \cite{Freund:2002ux} and on inclusive charm production \cite{magvicprl}. 

\begin{figure}[h]
\centering
\label{figarmesto}
\includegraphics[scale=0.5]{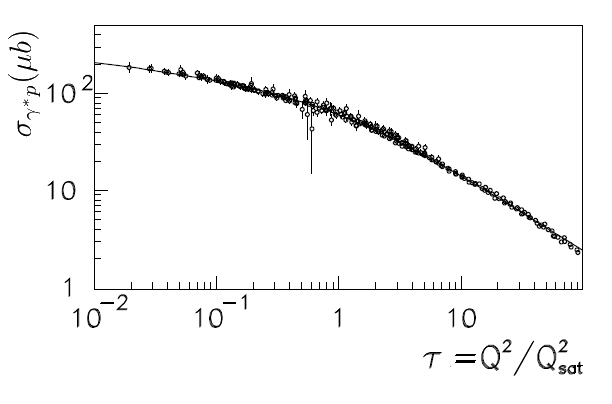}
\caption{Geometrical scaling for $\gamma^* p$ together with the scaling expression Eq. (\ref{sigtot_param_tau}). Figure from \cite{Armesto_scal}.}
\end{figure}

Concerning lepton-nucleus interactions, in Ref. \cite{Armesto_scal} the $\gamma^*A$ cross section is obtained from the corresponding cross section for $\gamma^*p$ process in the form 
\begin{eqnarray}
 \sigma^{\gamma^*A}_{tot}\,(\tau_A)  =  \frac{\pi R_A^2}{\pi R_p^2}\,\sigma^{\gamma^*p}_{tot}\,\left(\tau_p\left[\frac{ \pi R_A^2}{A\pi R_p^2}\right]^{\Delta}\right),
\label{nuclear_scaling}
\end{eqnarray}
where $\tau_p = Q^2/Q_{sat}^2$ is the saturation scale for a proton target.  The nuclear saturation scale was assumed to rise with the quotient of the transverse parton densities to the power $\Delta $. The nucleon saturation momentum is set to be $Q^2_{sat}=x_0/x^{\lambda}GeV^2$, where $x_0= 3.04\times 10^{-4}$, $\lambda=0.288$ and $\bar{x}=x/[1+ (4m_f^2/Q^2)]$, with $m_f=0.14 GeV$, as taken from the usual Golec Biernat-W\"usthoff model \cite{GBW}. The nuclear radius is given by $R_A=(1.12 A^{1/3}-0.86 A^{-1/3})$ fm. The following scaling curve for the photoabsortion cross section was considered \cite{Armesto_scal}:
\begin{eqnarray}
  \sigma^{\gamma^* p}_{tot}\,(\tau_p) = \bar{\sigma}_0\,
  \left[ \gamma_E + \Gamma\left(0,\nu \right) +
         \ln \left(\nu \right) \right],
 \label{sigtot_param_tau}
\end{eqnarray}
where $\nu = a/\tau_p^{b}$, $\gamma_E$ is the Euler constant and $\Gamma\left(0,\nu \right)$
the incomplete Gamma function. The parameters for the proton case were obtained from a fit to the small-$x$ $ep$ DESY-HERA data, producing $a=1.868$, $b=0.746$ and the overall  normalization was fixed by $\bar\sigma_0=[40.56]\mu b$. Their fit is presented in Fig. \ref{figarmesto}. The parameters for the nuclear saturation scale were determined by fitting the available lepton-hadron data using the relation in Eq. (\ref{nuclear_scaling}) and the same scaling function, Eq. (\ref{sigtot_param_tau}). They obtained $\delta=1/\Delta = 0.79\pm0.02$ and $\pi R_p^2=[1.55 \pm 0.02]fm^2$. 

In Ref. \cite{MS} it was demonstrated that the data on diffractive DIS, $\gamma^*p\rightarrow Xp$, and other diffractive observables present geometric scaling on the variable $\tau_D = Q^2/Q_{sat}^2(x_{\pom})$, in region $x_{\pom}<0.01$, where $x_{\pom} = (Q^2+M_X^2)/(Q^2+W^2)$. Moreover, the total cross sections for $\rho$, $\phi$ and $J/\psi$ are shown to present scaling on the variable $\tau_V = (Q^2+M_V^2)/Q_{sat}^2(x_{\pom})$. What that means is that not only the total cross section is a function of a single variable, but the cross section for vector meson production is also a function of a single variable, and the data agrees with that.  Nevertheless, \cite{MS} provides no theoretical or phenomenological expression for the scaling function. In this work, we extend the approach presented in Ref. \cite{Armesto_scal}  to exclusive (diffractive) processes to describe also the observed scaling features demonstrated in Ref. \cite{MS}. Based on the eikonal model in impact parameter space, we provide an expression for the cross section for exclusive production of vector mesons and DVCS as well. This expression provides a reasonable description for the available data for $V=\rho,\,\phi,\, J/\psi$ and real photons. The results are improved  by allowing a global fit using the universal scaling expression which depends on very few parameters. These theoretical and phenomenological results have direct consequences on prediction for the future electron-ion colliders \cite{EICs} and also for vector meson photo-production measured in ultra-peripheral nucleus-nucleus collisions at the LHC \cite{Glauber1,Glauber2}.

\section{\label{sec:level2} Cross sections for exclusive vector meson production and DVCS}

In general, one may obtain the total cross section for a hadronic interaction from the elastic scattering amplitude $a(s,b)$ through the optical theorem. In terms of the impact parameter, one can write
\begin{eqnarray}
\sigma_{tot} & = & 2\int d^2b \,\mathrm{Im}\, a(s,b),\\
\sigma_{el} & = & \int d^2b \,|a(s,b)|^2.
\end{eqnarray}
In the eikonal approach, $a(s,b) = i(1-e^{-\Omega(s,b)})$, where the eikonal $\Omega$ is a real function. Thus, $P(s,b)=e^{-2\,\Omega(s,b)}$ gives the probability that no inelastic interaction takes place at impact parameter $b$. Assuming for simplicity a Gaussian form for the eikonal, $\Omega(s,b) = \nu(s)\exp\left( -b^2/R^2 \right)$, analytical expressions for total and elastic cross sections are generated,
\begin{eqnarray}
\label{sigtot}
\sigma_{tot} & = & 2\pi R^2\left[\ln(\nu) +\gamma_E+\Gamma\left(0,\nu \right) \right],\\
\sigma_{el} & = & \pi R^2\left[\ln \left(\frac{\nu}{2}\right) +\gamma_E-\Gamma\left(0,2\nu \right) + 2\,\Gamma\left(0,\nu \right)  \right].\label{sigel}
\end{eqnarray}
From this, one can easily identify that Eq. (\ref{sigtot_param_tau}) relies on the total cross section from the eikonal model, Eq. (\ref{sigtot}), with the following identification, $\bar{\sigma}_0 = 2\pi R^2$ and $\nu = a/\tau_p^b$. The $a$ and $b$ parameters absorb the lost information when using a oversimplified photon wave-function overlap $\Phi^{\gamma^*\gamma^*}\propto \delta \left(r-1/Q\right)$ within the color dipole framework. We then construct the scaling function for describing exclusive diffractive processes starting from Eq. (\ref{sigel}). The main point is to associate the exclusive vector meson production and DVCS process as a quasi-elastic scattering.

For vector meson production, we have to include information related to the meson wave-function and in the DVCS case information on the real photon appearing in the final state. Adding this new information will modify the overall normalization in Eq. (\ref{sigel}) and possibly also the parameter $a$ and $b$ considered in Ref. \cite{Armesto_scal}. From the dipole model, one in general has \cite{Kowalski:2006hc}
\begin{eqnarray}
\label{sigel2}
\sigma_{tot} \propto \alpha_{em}\sum_f e_f^2,
\end{eqnarray}
where $e_f$ is the charge of the quark of flavor $f$ in the dipole picture, summed over all flavors. Also, for vector meson production
\begin{eqnarray}
\label{sigvm}
\sigma_{VM} \propto 4\pi \alpha_{em} \hat{e}_f^2 \frac{f_v^2}{M_V^2},
\end{eqnarray}
where $f_V$ is the meson coupling to the electromagnetic current and $\hat{e}_f$ is the effective charge of the quark in the meson. We then write
\begin{eqnarray}
\sigma (\gamma^*p\rightarrow Ep)  =  \frac{\bar{\sigma}_E}{2}\left[\ln \left(\frac{\nu}{2}\right) +\gamma_E-\Gamma\left(0,2\nu \right) + 2\,\Gamma\left(0,\nu \right)  \right],\label{sigexcl}
\end{eqnarray}
where $\bar{\sigma}_E = \bar{\sigma}_V$ in case of vector mesons and $\bar{\sigma}_E = \bar{\sigma}_{\mathrm{DVCS}}$  for DVCS process. In both cases, $\nu = a/\tau^b$, with $\tau = (Q^2 + M_V^2)/Q^2_\mathrm{sat}$ for exclusive production of mesons and $\tau = Q^2/Q^2_\mathrm{sat}$ for DVCS. Explicitly, the overall normalization of cross sections is obtained from the inspection of the overlap functions in Eqs. (\ref{sigel2}) and (\ref{sigvm}). Therefore, the final expressions for the overall normalization in our scaling function are given by
\begin{eqnarray}
\label{sig0dvcs}
\bar{\sigma}_{\mathrm{DVCS}} & = & \left(\alpha_e\,\sum_f e_f^2  \right)\bar{\sigma}_0, \\
\label{sig0v}
\bar{\sigma}_V & = & \frac{4\pi \hat{e}^2_ff_V^2}{M_V^2\left(\sum_f e_f^2 \right)}\bar{\sigma}_0.
\end{eqnarray}

\section{\label{sec:level3} Results}

\begin{figure}
\centering
\label{fig:res1}
\includegraphics[scale=0.35]{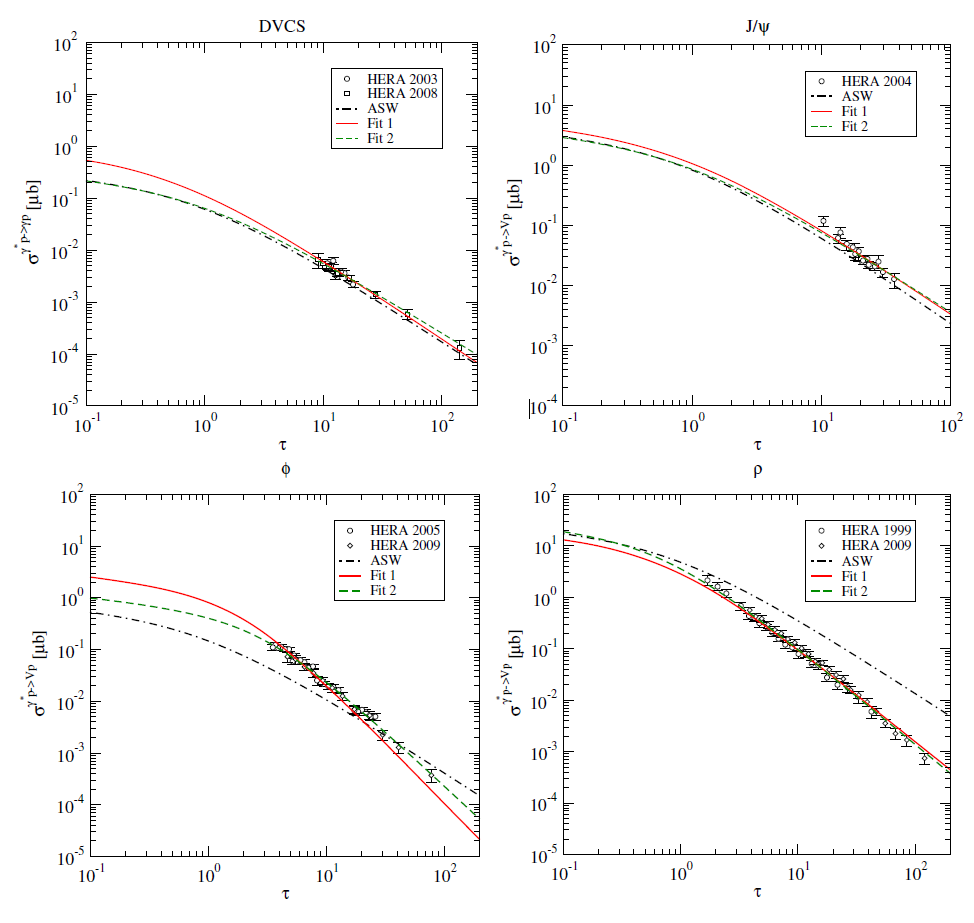}
\caption{The cross sections for DVCS, $\phi$, $J/\psi$ and $\rho$ as a function of $\tau$. Figure from \cite{ben2017}.}
\end{figure}

Let us now compare the scaling curve, Eq. (\ref{sigexcl}), to the available experimental data in small-$x$ lepton-proton collisions. The data sets we have considered are presented in Refs. \cite{dvcs,rho,phi,jpsi}. The values of parameters $M_V$, $f_V$ and $\hat{e}_V$ were taken  from Ref. \cite{Kowalski:2006hc}. We perform a fit to the experimental data using MINPACK routines~\cite{minpack} for choices of sets of parameters, described in the following. Our results are presented in Figures 2 and 3. Explicitly, the scaling variable is $\tau = \tau_V = (Q^2 + M_V^2)/Q^2_\mathrm{sat}(x)$ for exclusive production of mesons and $\tau = Q^2/Q^2_\mathrm{sat}(x)$ for DVCS, with $Q^2_{\mathrm{sat}}(x)=[(x_0/\bar{x})^{\lambda}]{GeV^2}$ as discussed in the introduction section.

We use two different choices to perform the fits. The first one, labeled ``Fit 1" in the figures, adjusts all the three parameters ($a$, $b$ and $\bar{\sigma}_0$). The other one, labeled ``Fit 2" in the figures, fits $a$, $b$ parameters, maintaining fixed $\bar{\sigma}_0 = [40.56]{\mu b}$. In general both fits describe in good agreement the available data for all observables (with the exception of $\phi$ meson) for photon-proton interactions. It is very clear that the quality of fit for Fit 1 and Fit 2 are somewhat equivalent. Fit 2 is a straightforward extension of the celebrated scaling curve presented in Ref. \cite{Armesto_scal} for the inclusive case. The overall normalization $\bar{\sigma}_0$ is common to inclusive and exclusive photon-target processes. For the sake of completeness, we also include the result using the original values for the parameters from the fitting to inclusive data \cite{Armesto_scal} (labeled by ASW in the curves) .

The geometrical scaling present in the lepton-proton cross sections for exclusive processes, as quantified by Eq. (\ref{sigexcl}),  is translated to the scattering on nuclear targets at high energies. Following the same arguments given in Ref. \cite{Armesto_scal}, the atomic number dependence is absorbed in the nuclear saturation scale and on the overall normalization related to the nuclear radius. Therefore, the cross section for lepton-nuclei scattering takes the following form, \begin{eqnarray}
 \sigma^{\gamma^*A\rightarrow EA}\,(\tau_A)  =  \frac{\pi R_A^2}{\pi R_p^2}\,\sigma^{\gamma^*p\rightarrow Ep}\,\left(\tau = \tau_A\right),
\label{nuclear_scaling2}
\end{eqnarray}
where the scaling variable in nuclear case is $\tau_A = \tau_p[\pi R_A^2/(A \pi R_p^2)]^{\Delta}$. In particular, we expect that for large $\tau_A$ the relation  is $\sigma (\gamma^*A\rightarrow EA)\propto R_A^2\,\tau_A^{-b}=R_A^2 \tau_p^{-b}(A^{1/3})^{\frac{b}{\delta}}$.  Figure 3 shows the cross sections for nuclear production of $J/\psi$ and $\rho$ as a function of the corresponding photon-nucleus energy, together with a plot of the prediction from Eq. (\ref{nuclear_scaling2}), using the paremeters adjusted for $\gamma^*p$ collisions. As the current data on nuclear targets are quite scarce at small-$x$ region, the scaling formula above can be tested in future measurements in EICs or in ultraperipheral heavy ions collisions. The robustness of the geometric scaling treatment for the interaction is quite impressive and similar scaling properties have been proved theoretically and experimentaly, for instance in charged hadron production \cite{prazlach} and in prompt photon production \cite{prazlapf} on $pA$ and $AA$ collisions in colliders energy regime.

\begin{figure}[h]
\centering
\label{res2}
\includegraphics[scale=0.35]{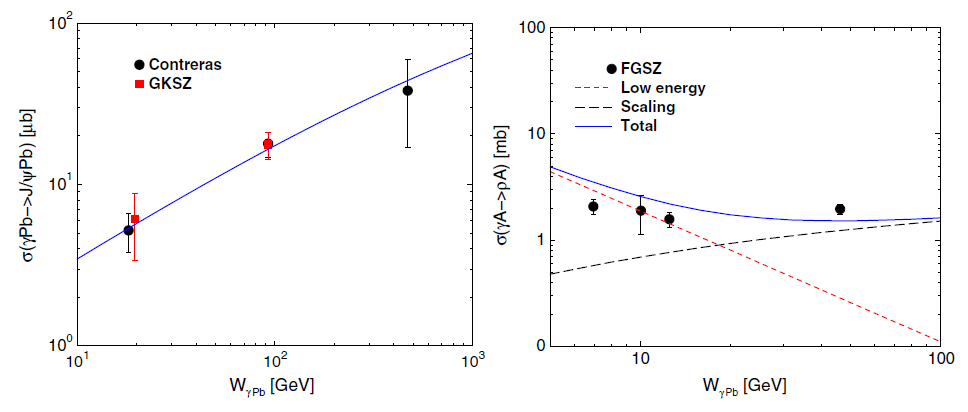}
\caption{The cross sections for nuclear production of $J/\psi$ and $\rho$ as a function of the corresponding photon-nucleus energy. Figure from \cite{ben2017}.}
\end{figure}

\section{Summary and conclusions \label{sec:sumcon}} 

This work demonstrates that by assuming geometric scaling phenomenon in exclusive processes at small-$x$ and simple considerations on the scope of  eikonal model, one is able to describe the available data on DVCS and vector meson production on nucleon target with a universal scaling function without any further parameter. We establish that the geometric scaling parametrization can be extrapolated to nuclear targets to be tested in future EICs or in ultra-peripheral collisions.  This implies that such dimensionless scale absorbs their energy and atomic number dependences. 

\begin{acknowledgments}

%\vspace{-0.4cm}

This work was financed by the Brazilian funding agency CNPq.

\end{acknowledgments}

\end{document}